\documentclass[11pt]{article}
\usepackage[totalwidth=460truept,totalheight=600truept]{geometry}
\usepackage{latexsym,amssymb,amsmath,graphicx}
\usepackage[T1]{fontenc}
\usepackage{color}

\global\arraycolsep=1truept
\newcommand{\beq}{\begin{equation}}
\newcommand{\eeq}{\end{equation}}
\def\bea{\begin{eqnarray}}
\def\eea{\end{eqnarray}}
\definecolor{darkred}{rgb}{.8,0,0}

\definecolor{darkblu}{rgb}{0,0,.8}

\makeatletter
\def\flE{\begin{picture}(0,0)
   \put( 0.25,    0){\vector( 1, 0){0.50}}
   \@ifstar{\@flE}{\@@flE}}
\def\@flE  #1{\put( 0.5 ,-0.03){\makebox(0,0)[ t]{$#1$}}\end{picture}}
\def\@@flE #1{\put( 0.5 , 0.03){\makebox(0,0)[ b]{$#1$}}\end{picture}}
\def\flNE{\begin{picture}(0,0)
   \put( 0.18, 0.18){\vector( 1, 1){0.64}}
   \@ifstar{\@flNE}{\@@flNE}}
\def\@flNE #1{\put( 0.52, 0.48){\makebox(0,0)[tl]{$#1$}}\end{picture}}
\def\@@flNE#1{\put( 0.48, 0.52){\makebox(0,0)[br]{$#1$}}\end{picture}}
\def\flN{\begin{picture}(0,0)
   \put(    0, 0.20){\vector( 0, 1){0.60}}
   \@ifstar{\@flN}{\@@flN}}
\def\@flN  #1{\put( 0.03, 0.5 ){\makebox(0,0)[ l]{$#1$}}\end{picture}}
\def\@@flN #1{\put(-0.03, 0.5 ){\makebox(0,0)[ r]{$#1$}}\end{picture}}
\def\flNW{\begin{picture}(0,0)
   \put(-0.18, 0.18){\vector(-1, 1){0.64}}
   \@ifstar{\@flNW}{\@@flNW}}
\def\@flNW #1{\put(-0.48, 0.52){\makebox(0,0)[bl]{$#1$}}\end{picture}}
\def\@@flNW#1{\put(-0.52, 0.48){\makebox(0,0)[tr]{$#1$}}\end{picture}}
\def\flW{\begin{picture}(0,0)
   \put(-0.25,    0){\vector(-1, 0){0.50}}
   \@ifstar{\@flW}{\@@flW}}
\def\@flW  #1{\put(-0.5 , 0.03){\makebox(0,0)[ b]{$#1$}}\end{picture}}
\def\@@flW #1{\put(-0.5 ,-0.03){\makebox(0,0)[ t]{$#1$}}\end{picture}}
\def\flSW{\begin{picture}(0,0)
   \put(-0.18,-0.18){\vector(-1,-1){0.64}}
   \@ifstar{\@flSW}{\@@flSW}}
\def\@flSW #1{\put(-0.52,-0.48){\makebox(0,0)[br]{$#1$}}\end{picture}}
\def\@@flSW#1{\put(-0.48,-0.52){\makebox(0,0)[tl]{$#1$}}\end{picture}}
\def\flS{\begin{picture}(0,0)
   \put(    0,-0.2 ){\vector( 0,-1){0.60}}
   \@ifstar{\@flS}{\@@flS}}
\def\@flS  #1{\put(-0.03,-0.5 ){\makebox(0,0)[ r]{$#1$}}\end{picture}}
\def\@@flS #1{\put( 0.03,-0.5 ){\makebox(0,0)[ l]{$#1$}}\end{picture}}
\def\flSE{\begin{picture}(0,0)
   \put( 0.18,-0.18){\vector( 1,-1){0.64}}
   \@ifstar{\@flSE}{\@@flSE}}
\def\@flSE #1{\put( 0.48,-0.52){\makebox(0,0)[tr]{$#1$}}\end{picture}}
\def\@@flSE#1{\put( 0.52,-0.48){\makebox(0,0)[bl]{$#1$}}\end{picture}}
\def\capsa(#1,#2)#3{\put(#1,#2){\makebox(0,0){$#3$}}}
\def\indiag{\@ifnextchar [{\@indiag}{\@indiag[15ex]}}
\def\@indiag[#1](#2,#3){\begingroup
   \setlength{\unitlength}{#1}
   \medskip
   \begin{center}
   \begin{picture}(#2,#3)}
\def\exdiag{\end{picture}
   \end{center}
   \medskip
   \endgroup}
\makeatother

\begin{document}

%\hfill UB-ECM-PF-09/24

%\hfill UB-ICC-09/xx

\vskip 1.4truecm

\begin{center}
{\LARGE
%\scshape
\bf Consistent Levi Civita truncation uniquely characterizes the Lovelock Lagrangians
\par}

\vskip 1.5truecm

\textsl{Naresh Dadhich}

\textsl{IUCAA, Post Bag 4, Ganeshkhind, Pune 411 007, India}

{\footnotesize nkd@iucaa.ernet.in}

\vskip 1.5truecm

\textsl{Josep M. Pons}

\textit{DECM and ICC, Facultat de F\'{\i}sica, Universitat de Barcelona,\\ Diagonal 647, E-08028 Barcelona, 
Catalonia, Spain.}

\textit{\&}

\textsl{IUCAA, Post Bag 4, Ganeshkhind, Pune 411 007, India}

{\footnotesize pons@ecm.ub.es}

\vskip 2truecm

\textbf{Abstract}
\end{center}

\bigskip

{\small We define the concept of Levi Civita truncation for a Lagrangian in the Palatini formulation with an arbitrary connection, and show that its consistency uniquely identifies the Lovelock Lagrangians.}

\vskip 1truecm

\vfill\eject

%%%%%%%%%%%%%%%%%%%%%%%%%%%%%%%%%%%%%%%%%%%%%%%
\section{Introduction}
\setcounter{equation}{0}
%%%%%%%%%%%%%%%%%%%%%%%%%%%%%%%%%%%%%%%%%%%%%%%

In the Palatini formulation of gravity, the connection is treated in line with gauge field theory as an independent variable. There are therefore two formulations, one with the metric (the vielbein) and the connection being both independent variables, and the other with the metric (the vielbein) alone being the independent variable. Of course it is expected that both these formulations should yield the same equation of motion (EOM). The proof of the equivalence of the two formulations of general relativity has a long history beginning with Einstein himself in 1925 \cite{ein25}, where he had established the equivalence under the additional condition on the connection of the vanishing of the trace of the torsion tensor. Since then several authors have worked on this problem (see for instance,   \cite{hehl,stachel,Floreanini:1990kt,Percacci:1990wy}) but none has improved upon the Einstein condition. In the usual derivations of the equivalence, the symmetry of the connection, which means vanishing torsion, is always assumed in the metric-affine formulation, whereas for the vielbein-affine formulation is assumed the antisymmetry of the spin-connection, which means metricity. This was the situation until very recently when in the works \cite{Sotiriou:2009xt,Dadhich:2010xa}, the equivalence is established for an arbitrary connection without any condition on it. It turns out that the Einstein condition is indeed a gauge condition because the action admits a gauge symmetry --the well known projective transformation-- which in our understanding has for the first time been recognised as such in \cite{Dadhich:2010xa}. Results in the same direction can be found in \cite{gia}.

\vspace{4mm}

Once we realize that it is possible to work with an arbitrary connection, with no restrictions whatsoever, it is pertinent to consider the extension of this result to  higher order gravity for a polynomial action. Using a symmetric connection in the metric-affine formulation, it has been argued that the equivalence of the two formulations identifies the Lovelock polynomial action uniquely \cite{Exirifard:2007da} (See also \cite{Exirifard:2009qj,Borunda:2008kf,BasteroGil:2009cn}). Although we believe that the result is true, however it appears that the arguments are not completely satisfactory. This is however an interesting characterization of the Lovelock Lagrangians. We would therefore like to address this question in a more general setting of an arbitrary connection. It turns out that it is very difficult to tackle this question in all generality. However, in the process, we have discovered a neat characterization of the Lovelock Lagrangians. This is precisely what we wish to demonstrate in this paper. 

We shall consider Lagrangians which are polynomial in the Riemann curvature without its derivatives. We employ the concept of truncation of the Palatini Lagrangian with respect to the Levi Civita (LC) connection by which we mean that the connection is substituted by the LC connection. The consistent LC truncation would mean that the truncated EOM of the Palatini Lagrangian is the same as the EOM of the truncated Lagrangian. Our main result is that the consistency of such a truncation uniquely identifies the Lovelock polynomial Lagrangians.

\vspace{4mm}

The paper is organized as follows: The next section would be devoted to the consistent LC truncation of the Palatini Lagrangian which would be followed by the establishement of the main result as a theorem. We conclude with a discussion.
%%%%%%%%%%%%%%%%%%%%%%%%%%%%%%%%%%%%%%%%%%%%%%%
\section{Consistent Levi Civita truncation}
\setcounter{equation}{0}
%%%%%%%%%%%%%%%%%%%%%%%%%%%%%%%%%%%%%%%%%%%%%%%

All through, we shall consider a Palatini Lagrangian which is polynomial in the Riemann curvature but free of its derivatives and the connection is arbitrary without any symmetry properties. This is quite different from the previous works where either the connection is usually taken as torsionless, symmetric connection in the metric-affine formalism or as metric compatible, antisymmetric in its flat indices for the vielbein-affine formalism 

We define the truncation \cite{Pons:2004ky,Pons:2009ch} as the introduction of constraints that eliminate some fields or field components either in the Lagrangian or in the EOM. The truncation is said to be consistent if the truncation of the EOM of the original Lagrangian agrees with the EOM of the truncated Lagrangian\footnote{There is in the literature a weaker version of the concept of consistent truncation, see for instance \cite{Duff:1985jd}. In this weaker sense, a truncation --setting to zero some fields-- is consistent if the solution of the EOM for the remaining fields can be uplifted to a solution of the original EOM. Now the truncated Lagrangian plays no role.}. We shall however not consider other type of truncations like those associated with the Kaluza-Klein dimensional reductions. Our interest is in the LC truncation where the arbitrary connection is replaced by the LC connection. We shall seek the Lagrangians satisfying the consistency condition.  

We shall write $\bar{\cal L}\rightarrow {\cal L}$ to denote the LC truncation. The overhead bar is indicative of the connection being an arbitrary independent field and the quantities built from it. Dropping the bar will indicate the LC truncation has been effected. The consistency of the truncation is indicated by the commutativity of the diagram

\indiag(1,1) \capsa(0,0){
\overline{EOM} } \capsa(0,1){\bar{\cal L}}
\capsa(0,1){\flS{{}}} \capsa(0,0){\qquad\flE{{\rm Trunc}}}
\capsa(0,1){\qquad\flE{{\rm Trunc}}} \capsa(1,0){\qquad\
\qquad  EOM  } \capsa(1.3,1){{\cal L}}
\capsa(1.3,1){\flS{{}}} 
\exdiag

This consistency guarantees that any solution of ${\cal L}$ can be uplifted to a solution of $\bar{\cal L}$\,. 

\vspace{4mm}

Let us start with some notation for the vielbein-affine formalism. The vielbein one forms are $e^{{}_{{{}^{I }}}}= e_\mu^{{}_{{{}^{I }}}} d x^\mu$; the two-form Riemann tensor is defined as $\bar R^{{}_{{}^{{}^{I J}}}}= d\,\bar\omega^{{}_{{}^{{}^{I J}}}} + \bar\omega^{{}_{{{}^{I }}}}_{{}_{{}^{{}^{\ K}}}}\wedge\, \bar\omega^{{}_{{}^{{}^{K J}}}}$, where $\bar\omega^{{}_{{}^{{}^{I J}}}}$ is the one-form connection, completely arbitrary. Its relation to the connection in the metric-affine formulation is $
\bar\omega_{\mu \  J}^{\ \, I} =e^I_\nu(\partial_\mu e^\nu_J + \bar \Gamma_{\  
\mu\rho}^\nu e^\rho_J)\,
$. It is convenient to perform a change of variables, by making use of the LC connection $\omega^{{}_{{}^{{}^{I J}}}}$ (torsionless and metric compatible). We define the one-form $C^{{}_{{}^{{}^{I J}}}}$ by 
\beq
\bar\omega^{{}_{{}^{{}^{I J}}}} = \omega^{{}_{{}^{{}^{I J}}}} + C^{{}_{{}^{{}^{I J}}}}\,.
\label{thechange}
\eeq
Under this change of variables, $\bar R^{{}_{{}^{{}^{I J}}}}$ becomes
\beq
\bar R^{{}_{{}^{{}^{I J}}}}= R^{{}_{{}^{{}^{I J}}}} + D\,C^{{}_{{}^{{}^{I J}}}} + C^{{}_{{{}^{I }}}}_{{}_{{}^{{}^{\ K}}}}\wedge\, C^{{}_{{}^{{}^{K J}}}}\,,
\label{thechangeR}
\eeq
where $R^{{}_{{}^{{}^{I J}}}}$ is the two-form Riemann tensor obtained from the LC connection and $D$ is the covariant differential operator associated with the LC connection
\beq
D\,C^{{}_{{}^{{}^{I J}}}} = d\,C^{{}_{{}^{{}^{I J}}}} + \omega^{{}_{{{I }}}}_{{}_{{}^{{}^{\ K}}}}\wedge\, C^{{}_{{}^{{}^{K J}}}}+ \omega^{{}_{{{J }}}}_{{}_{{}^{{}^{\ K}}}}\wedge\, C^{{}_{{}^{{}^{I K}}}}
= \nabla_{\!\mu} C^{{}_{{}^{{I J}}}}_\nu dx^\mu\wedge dx^\nu\,,
\label{covdiff}
\eeq
where $\nabla_{\!\mu}$ is the LC covariant derivative.

\vspace{4mm}

Once the change of variables is performed, the LC truncation is defined by setting $C=0$. But before the truncation, the Lagrangian $\bar{\cal L}$ is a functional of the vielbein and the $C$ tensor. In fact, the substitution (\ref{thechangeR}) yields
\beq
\bar{\cal L}={\cal L} + {\cal O}(C),
\label{barl-c}
\eeq
where ${\cal L}$ is the LC truncated Lagrangian and by ${\cal O}(C)$ we mean terms at least linear in $C$ and including its first derivatives.

It may be noted that if the term linear in $C$ cannot be absorbed in the total divergence, the consistency of the truncation would be a non-trivial condition. 

%%%%%%%%%%%%%%%%%%%%%%%%%%%%%%%%%%%%%%%%%%%%%%%
\section{Lovelock characterization}
\setcounter{equation}{0}
%%%%%%%%%%%%%%%%%%%%%%%%%%%%%%%%%%%%%%%%%%%%%%%

To analyze the consistency of the truncation, 
we must compare the truncated EOM with the EOM of the truncated Lagrangian. With this purpose in mind, and as a matter of convenience, let us express the Riemann tensor in the holonomic, coordinate basis (although our framework continues to be the formalism with vielbein). In our conventions
$\displaystyle R_{\mu\nu  \ J}^{\ \ \ I} = - e^\rho_J e^I_\sigma R_{\mu\nu\rho}^{\ \ \ \ \sigma}$\,,
and the truncated Lagrangian will be written as a function of the vielbein and the Riemann tensor in the form $\ {\cal L}(e^I_\lambda, R_{\mu\nu\rho\sigma}[g,\Gamma[g]])$. 

Since in the truncated theory the Riemann tensor is built with the LC connection, keeping all its indices down will allow us to take full advantage of the symmetry properties of $R_{\mu\nu\rho\sigma}$ which will be inherited as symmetry properties by the density tensor $\displaystyle\frac{\partial {\cal L}}{\partial R_{\mu\nu\rho\sigma}}$. 

We now have the two sides of the truncation as follows:

\vspace{4mm}

{\bf A: LC truncated EOM from the Lagrangian $\bar{\cal L}$}
\beq
\frac{\partial {\cal L}}{\partial e^I_\mu} + \Big(\frac{\partial {\cal L}}{\partial R_{\alpha\beta\rho\mu}} R_{\alpha\beta\rho}^{\ \ \ \ \lambda} +\frac{\partial {\cal L}}{\partial R_{\alpha\beta\rho\lambda}} R_{\alpha\beta\rho}^{\ \ \ \ \mu}\Big)e_{{}_{{}^{\lambda \,I}}}=0\,,
\label{eomsimpl}
\eeq
\beq
\nabla_{\!\mu}\frac{\partial {\cal L}}{\partial R_{\mu\nu\rho\sigma}}=0\,,
\label{eomextra}
\eeq

\vspace{4mm}

{\bf B: EOM from the LC truncated Lagrangian ${\cal L}$}
\beq
\frac{\delta {\cal L}}{\delta e^I_\mu} = \frac{\partial {\cal L}}{\partial e^I_\mu} + \Big(\frac{\partial {\cal L}}{\partial R_{\alpha\beta\rho\mu}} R_{\alpha\beta\rho}^{\ \ \ \ \lambda} +\frac{\partial {\cal L}}{\partial R_{\alpha\beta\rho\lambda}} R_{\alpha\beta\rho}^{\ \ \ \ \mu} + 4 \nabla_{\!\rho}\nabla_{\!\alpha}\frac{\partial {\cal L}}{\partial R_{\alpha(\mu\lambda)\rho}}\Big) e_{{}_{{}^{\lambda \,I}}}=0\,.
\label{eomtrunc}
\eeq

\vspace{4mm}

Let us note that Eq. (\ref{eomextra}) originates from the term linear in $C$ in Eq. (\ref{barl-c}),
\beq
\frac{\delta\bar {\cal L}}{\delta C}\vert_{C=0}\ \Leftrightarrow\ \nabla_{\!\mu}\frac{\partial {\cal L}}{\partial R_{\mu\nu\rho\sigma}}=0\,.
\label{eomextra2}
\eeq
In the absence of such a linear term, modulo a divergence, Eq. (\ref{eomextra}) will be an identity and vice-versa.

\vspace{4mm}

It is obvious that {\bf A}$\Rightarrow${\bf B}. The LC truncation will be consistent if the other way around is also true. Our main result is then the following theorem.  

 \vspace{4mm}

\underline{\bf Theorem:}

{\sl The Levi-Civita truncation is consistent if and only if the Lagrangian, which is a polynomial in the Riemann curvature and not involving its derivatives, is the well-known Lovelock Lagrangian.}

 \vspace{4mm}

It could be also cast in a larger set of equivalent statements as follows: 
%The cornerstone to this result is the analysis of equation (\ref{eomextra}). In fact the previous theorem can be cast within a larger set of equivalent statements, as follows}

{\sl 
\vspace{3mm}

{\bf I}. Modulo a divergence, $\bar{\cal L}$ is at least quadratic in $C$,

\vspace{3mm}

{\bf II}.
The LC truncation (setting $C = 0$), $\ \bar{\cal L}\rightarrow {\cal L}\ $, is consistent,

\vspace{3mm}

{\bf III}.
The EOM for the LC truncated Lagrangian ${\cal L}$ is of second order,

\vspace{3mm}

{\bf IV}.
The expression $\displaystyle \nabla_{\!\mu}\frac{\partial {\cal L}}{\partial R_{\mu\nu\rho\sigma}}$ vanishes identically,

\vspace{3mm}

{\bf V}.
The truncated Lagrangian ${\cal L}$ is a Lovelock polynomial.
}
\vspace{4mm}

Note that the Theorem corresponds to the equivalence: ${\bf II}\Leftrightarrow {\bf V}$. The cornerstone of our analysis is the equation (\ref{eomextra}).

\vspace{4mm}

{\bf \underline{Proof}}

\vspace{4mm}

The equivalence ${\bf I}\Leftrightarrow {\bf IV}$ has already been shown after Eq. (\ref{eomextra2}). Now we will prove ${\bf II}\Leftrightarrow {\bf IV}$, i.e, that the necessary and sufficient condition for the LC truncation to be consistent is that Eq. (\ref{eomextra}) becomes an identity. The sufficiency is quite obvious and we need only to prove its necessity.

To prove this, let us consider the case when Eq. (\ref{eomextra}) is non-empty. The Lagrangian ${\cal L}$ is a sum of homogeneous polynomials, we need only to consider a homogeneous polynomial. The general form of the LC truncated Lagrangian ${\cal L}$ is
\beq
{\cal L}= R_{\mu_{1}\nu_{1}\rho_{1}\sigma_{1}}...R_{\mu_{k}\nu_{k}\rho_{k}\sigma_{k}}\, A^{\mu_{1}\nu_{1}\rho_{1}\sigma_{1}...\mu_{k}\nu_{k}\rho_{k}\sigma_{k}}\,,
\label{generalform}
\eeq
with the coefficient $\displaystyle A^{\mu_{1}\nu_{1}\rho_{1}\sigma_{1}...\mu_{k}\nu_{k}\rho_{k}\sigma_{k}}$, made exclusively with the metric $g^{\mu\nu}$ (the vielbein having been absorbed), exhibiting all the symmetries of the Riemann tensor and in addition the symmetry of exchange of $\mu_{i}\nu_{i}\rho_{i}\sigma_{i}$ and $\mu_{j}\nu_{j}\rho_{j}\sigma_{j}$. 
Then the expression 
$
\displaystyle \nabla_{\!\mu}\frac{\partial {\cal L}}{\partial R_{\mu\nu\rho\sigma}}\ 
$
will read as  
\beq
\displaystyle \nabla_{\!\mu}\frac{\partial {\cal L}}{\partial R_{\mu\nu\rho\sigma}}=k(k-1)\, (\nabla_{\!\mu}R_{\mu_{2}\nu_{2}\rho_{2}\sigma_{2}})R_{\mu_{3}\nu_{3}\rho_{3}\sigma_{3}}...R_{\mu_{k}\nu_{k}\rho_{k}\sigma_{k}}\, A^{\mu\nu\rho\sigma\mu_{2}\nu_{2}\rho_{2}\sigma_{2}...\mu_{k}\nu_{k}\rho_{k}\sigma_{k}}\,.
\label{keystone-dev}
\eeq
Since this does not vanish, the symmetries of $A^{\mu\nu...}$ do not allow the Bianchi identity to operate on the derivative term on the right. Now let us apply another covariant derivative to Eq. (\ref{keystone-dev})
\bea
\displaystyle && \nabla_{\!\rho}\nabla_{\!\mu}\frac{\partial {\cal L}}{\partial R_{\mu\nu\rho\sigma}}=k(k-1)\, (\nabla_{\!\rho}\nabla_{\!\mu}R_{\mu_{2}\nu_{2}\rho_{2}\sigma_{2}})R_{\mu_{3}\nu_{3}\rho_{3}\sigma_{3}}...R_{\mu_{k}\nu_{k}\rho_{k}\sigma_{k}}\, A^{\mu\nu\rho\sigma\mu_{2}\nu_{2}\rho_{2}\sigma_{2}...\mu_{k}\nu_{k}\rho_{k}\sigma_{k}}
\nonumber\\
&+& k(k-1)(k-2)\, (\nabla_{\!\mu}R_{\mu_{2}\nu_{2}\rho_{2}\sigma_{2}})(\nabla_{\!\rho}R_{\mu_{3}\nu_{3}\rho_{3}\sigma_{3}})R_{\mu_{4}\nu_{4}\rho_{4}\sigma_{4}}...R_{\mu_{k}\nu_{k}\rho_{k}\sigma_{k}}\, A^{\mu\nu\rho\sigma\mu_{2}\nu_{2}\rho_{2}\sigma_{2}...\mu_{k}\nu_{k}\rho_{k}\sigma_{k}}
\,
\label{keystone-dev2}
\eea
So we would have the term, $\displaystyle \partial_{\!\rho}\partial_{\!\mu}R_{\mu_{2}\nu_{2}\rho_{2}\sigma_{2}}$ which would certainly be fourth order in the derivatives unless there is antisymmetry in $\rho$ and $\mu$. In that case the Riemann tensor in Eq. (\ref{generalform}) will appear antisymmetrized with respect to all their four indices leading to the vanishing of the Lagrangian itself because of the non-differential Bianchi identity, $\displaystyle R_{[\mu\nu\rho]\sigma} = 0$. We are thus led to the conclusion that Eq. (\ref{keystone-dev2}) would be fourth order in the derivatives for non-empty Eq. (\ref{eomextra}). This means Eq. 
(\ref{eomtrunc}) would also be fourth order \cite{footnot}. Thus  we conclude that 
when Eq. (\ref{eomextra}) is non-empty, Eq. (\ref{eomtrunc}) is a fourth order EOM which cannot be equivalent to second order Eq. (\ref{eomsimpl}) plus third order Eq. (\ref{eomextra}).  This shows that the LC truncation is not consistent and we have therefore proved the necessary condition and so ${\bf II}\Leftrightarrow {\bf IV}$. By the same token ${\bf III}\Leftrightarrow {\bf IV}$ is also proved because for Eq. (\ref{eomtrunc}) to be of second order we have just shown that it is necessary (and obviously sufficient) that Eq. (\ref{eomextra}) be an identically vanishing equation.

\vspace{4mm}

To finish the proof of the theorem we need to show that Eq. (\ref{keystone-dev}) admits only the Lovelock Lagrangians as solutions. We look for what structures of $A^{\mu\nu...}$ make Eq. (\ref{keystone-dev}) to vanish identically. This can only happen if the Bianchi identity operates in Eq. (\ref{keystone-dev}), 
and for that to occur additional antisymmetry of the indices of $A^{\mu\nu...}$ would be required. This Bianchi identity can involve the two first indices of  $\displaystyle R_{\mu\nu\rho\sigma}$ or the last two indices. We can use the symmetries of the Riemann tensor to assume that the first two indices are those involved in the Bianchi identity. This means that $A^{\mu\nu...}$ in Eq. (\ref{generalform}) has the extra antisymmetrization for the indices $\mu_i,\mu_j$. Since there was already the antisymmetrization $\mu_i,\nu_i$, we end up with all the $\mu_i,\nu_j$ indices antisymmetrized. On the other hand, the symmetries of the Riemann tensor imply that one can use any of the four indices of the Riemann tensor in the covariant derivative present in Eq. (\ref{keystone-dev}). Using for instance $\displaystyle \nabla_{\!\rho}\frac{\partial {\cal L}}{\partial R_{\mu\nu\rho\sigma}}$, the same arguments used above show that there must be an antisymmetrization for all the $\rho_i,\sigma_j$ indices as well.
With this additional symmetry, let us rewrite Eq. (\ref{generalform}) as 
\beq
{\cal L}= R_{\mu_{1}\nu_{1}}^{\quad\ \rho_{1}\sigma_{1}}...R_{\mu_{k}\nu_{k}}^{\quad\ \rho_{k}\sigma_{k}}\, A^{\mu_{1}\nu_{1}...\mu_{k}\nu_{k}}_{\qquad\qquad\rho_{1}\sigma_{1}...\rho_{k}\sigma_{k}}\,,
\label{generalform2}
\eeq
with $A^{\mu\nu...}$ antisymmetric in all the upper and all the lower indices.
Up to a global factor (determined by the scalar density requirement for the Lagrangian), the only geometric structure that can accord to this requirement is 
\beq
\delta^{\mu_1\dots \mu_r}_{\nu_1\dots \nu_r} = \delta^{[\mu_1}_{\nu_1}\dots\delta^{\mu_r]}_{\nu_r}
= \delta^{\mu_1}_{[\nu_1}\dots\delta^{\mu_r}_{\nu_r]}\,.
\label{multiK}
\eeq
So we end up with a Lovelock Lagrangian \cite{Lovelock:1971yv}
\beq\label{lovelocklag}
 {\cal L}_p= \sqrt{-g}\, \delta^{\mu_1\dots \mu_{p}\nu_1\dots \nu_p}_{\rho_1\dots \rho_{p}\sigma_1\dots \sigma_p} \, 
 R_{\mu_1\nu_1} \!{}^{\!\rho_1\sigma_1}\dots  R_{\mu_p\nu_p}\! {}^{\!\rho_p \sigma_p}\,.
\eeq
We have thus proved that the general solution of Eq. (\ref{eomextra}) vanishing identically is ${\cal L} = \sum_{p} c_p {\cal L}_p$ with arbitrary coefficients $c_p$ (The fact that the Lovelock Lagrangians have Eq. (\ref{eomextra}) vanishing identically has already been shown in \cite{Mukhopadhyay:2006vu,Yale:2010jy}).

This proves ${\bf IV}\Leftrightarrow {\bf V}$ and establishes the equivalence of all the statements. Consequently it proves the theorem.

\vspace{4mm}

Alternatively \cite{Myers:1987yn,jorgez2000}, we can also write ${\cal L}_p$ as
\beq
{\cal L}_p = \epsilon_{{}_{I_1 I_2 ... I_n}}R^{{}_{{}^{{}^{I_1 I_2}}}}\wedge...\wedge R^{{}_{{}^{{}^{I_{2p-1} I_{2p}}}}}\wedge e^{2p+1}\wedge...\wedge e^{I_n}\,
\eeq
where $n$ is the spacetime dimension. This is the LC truncated Lagrangian obtained by dropping the bars from the original Lovelock-Palatini Lagrangian. As recognised earlier for the Einstein-Palatini Lagrangian in \cite{Dadhich:2010xa}, the same gauge symmetry  
\beq
 \bar\omega^{{}_{{}^{{}^{I J}}}}\rightarrow \bar\omega^{{}_{{}^{{}^{I J}}}}+
 \eta^{{}_{{}^{{}^{I J}}}}\,V\,,
\label{thegauges}
\eeq 
with $V$ an arbitrary one-form, is admitted by this Lagrangian (in the metric affine formulation, the gauge symmetry is $\displaystyle \bar \Gamma_{\  
\mu\rho}^\nu\rightarrow\bar \Gamma_{\  
\mu\rho}^\nu+V_\mu \delta^\nu_\rho$). It can be seen easily that $\bar{\cal L}_{p}$ is invariant because  
\beq
\bar R^{{}_{{}^{{}^{I J}}}} \rightarrow \bar R^{{}_{{}^{{}^{I J}}}} + \eta^{{}_{{}^{{}^{I J}}}}\, d\,V.
\label{thegaugesR}
\eeq

%%%%%%%%%%%%%%%%%%%%%%%%%%%%%%%%%%%%%%%%%%%%%%%
\section{Discussion}
%%%%%%%%%%%%%%%%%%%%%%%%%%%%%%%%%%%%%%%%%%%%%%%

Our analysis is critically based on the equation (\ref{eomextra}) being an identity. It is a necessary and sufficient condition for the LC truncation to be consistent as well as the EOM being second order and the Lagrangian being the Lovelock. It may be noted that all previous considerations of this problem referred to a symmetric connection in the metric affine formulation while here we deal with an arbitrary connection without any restrictions. This is a step forward. 

It is remarkable and interesting to note that the consistent LC truncation requirement for an arbitrary connection is precisely what identifies uniquely the Lovelock Lagrangians. The Lovelock Lagrangians are not only the most natural generalization of the Einstein-Hilbert Lagrangian but they also imbibe the physically most desirable feature of the EOM being second order quasi-linear. That allows the formulation of the initial value problem and thereby ensuring a unique evolution modulo diffeomorphism invariance for a given initial conditions. This suggests that the consistent LC truncation may at some deep level be linked to this important physical property. Above all it seems to provide some new insight. In this regard, it is also interesting to recognise yet another novel characterization of the Lovelock Lagrangians by the Bianchi derivative \cite{Dadhich:2008df}. In here, the idea of constructing a second rank divergence free differential operator from the vanishing trace of the Bianchi derivative of the Riemann curvature is extended to the entire Lovelock polynomials. 

Since we employ an arbitrary connection, a new gauge symmetry, Eq. (\ref{thegauges}), which was first noticed for the Einstein-Palatini Lagrangian \cite{Dadhich:2010xa}, is carried over to the Lovelock-Palatini Lagrangians as well. If the preservation of this gauge invariance plays a role in the selection of the physically relevant theories, the Lovelock-Palatini Lagrangians would pass the test.

We had begun with the more general question of the equivalence between the Palatini and the metric formulations and this equivalence was shown to exist for the Einstein-Palatini Lagrangian with an arbitrary connection \cite{Dadhich:2010xa}. This is because the general solution for the
$C$ tensor is a pure gauge solution. The question is to carry forward this equivalence for the Lovelock-Palatini case as well. However here we have not been able to establish the equivalence although the consistent LC truncation does indeed identify the Lovelock Lagrangians uniquely. This is what would engage us for some time to come.

%%%%%%%%%%%%%%%%%%%%%%%%%%%%%%%%%%%%
\section*{Aknowledgments}
%%%%%%%%%%%%%%%%%%%%%%%%%%%%%%%%%%%%
We thank Q. Exirifard, B. Janssen, B. Pitts and J. Zanelli for useful correspondence. JMP acknowledges support by MCYT FPA 2007-66665, CIRIT GC 2005SGR-00564 and Spanish Consolider-Ingenio 2010 Programme CPAN (CSD2007-00042). He also acknowledges IUCAA and its people for the warm hospitality.


\begin{thebibliography}{99}

\bibitem{ein25} 
A.~Einstein, ``Einheitliche Fieldtheorie von Gravitation und Elektrizit\"at``, Pruess. Akad. Wiss. {\sl 414}, 1925; A.~Unzicker and T.~Case,
``Translation of Einstein's attempt of a unified field theory with teleparallelism,''arXiv:physics/0503046.

%\cite{hehl}
\bibitem{hehl}
F.~W.~Hehl and G.~D.~Kerlick, 
``Metric-Affine Variational Principles in General 
Relativity. I. Riemannian Space-Time,''
Gen.\ Rel.\ Grav. {\bf 9} (1978) 691

\bibitem{stachel} 
A.~Papapetrou and J.~Stachel,
``A New Lagrangian for the Vacuum Einstein Equations and Its Tetrad Form,''
Gen.\ Rel.\ Grav. {\bf 9} (1978) 1075

%\cite{Floreanini:1990kt}
\bibitem{Floreanini:1990kt}
  R.~Floreanini and R.~Percacci,
  ``Palatini formalism and new canonical variables for GL(4) invariant
  gravity,''
  Class.\ Quant.\ Grav.\  {\bf 7} (1990) 1805.
  %%CITATION = CQGRD,7,1805;%%

%\cite{Percacci:1990wy}
\bibitem{Percacci:1990wy}
  R.~Percacci,
  ``The Higgs Phenomenon in Quantum Gravity,''
  Nucl.\ Phys.\  B {\bf 353} (1991) 271
  [arXiv:0712.3545 [hep-th]].
  %%CITATION = NUPHA,B353,271;%%

%\cite{Sotiriou:2009xt}
\bibitem{Sotiriou:2009xt}
T.~P.~Sotiriou,
``f(R) gravity, torsion and non-metricity,''
Class.\ Quant.\ Grav.\  {\bf 26} (2009) 152001
[arXiv:0904.2774 [gr-qc]].
%%CITATION = CQGRD,26,152001;%%

%\cite{Dadhich:2010xa}
\bibitem{Dadhich:2010xa}
  N.~Dadhich and J.~M.~Pons,
  ``Equivalence of the Einstein-Hilbert and the Einstein-Palatini formulations of general relativity for an 
arbitrary connection,''
  arXiv:1010.0869 [gr-qc].
  %%CITATION = ARXIV:1010.0869;%%

%\cite{gia}
\bibitem{gia}
  G.~Giachetta and L.~Mangiarotti,
  ``Projective invariance and Einstein Equations,''
  arXiv:1010.0869 [gr-qc].


%\cite{Exirifard:2007da}
\bibitem{Exirifard:2007da}
  Q.~Exirifard and M.~M.~Sheikh-Jabbari,
  ``Lovelock Gravity at the Crossroads of Palatini and Metric Formulations,''
  Phys.\ Lett.\  B {\bf 661} (2008) 158
  [arXiv:0705.1879 [hep-th]].
  %%CITATION = PHLTA,B661,158;%%

%\cite{Exirifard:2009qj}
\bibitem{Exirifard:2009qj}
  Q.~Exirifard,
  ``Generalized Lovelock gravity,''
  arXiv:0911.2872 [gr-qc].
  %%CITATION = ARXIV:0911.2872;%%

%\cite{Borunda:2008kf}
\bibitem{Borunda:2008kf}
  M.~Borunda, B.~Janssen and M.~Bastero-Gil,
  ``Palatini versus metric formulation in higher curvature gravity,''
  JCAP {\bf 0811} (2008) 008
  [arXiv:0804.4440 [hep-th]].
  %%CITATION = JCAPA,0811,008;%%

%\cite{BasteroGil:2009cn}
\bibitem{BasteroGil:2009cn}
  M.~Bastero-Gil, M.~Borunda and B.~Janssen,
  ``The Palatini formalism for higher-curvature gravity theories,''
  AIP Conf.\ Proc.\  {\bf 1122} (2009) 189
  [arXiv:0901.1590 [hep-th]].
  %%CITATION = APCPC,1122,189;%%

%\cite{Pons:2004ky}
\bibitem{Pons:2004ky}
  J.~M.~Pons and P.~Talavera,
  ``Truncations driven by constraints: Consistency and conditions for correct
  upliftings,''
  Nucl.\ Phys.\  B {\bf 703}, 537 (2004)
  [arXiv:hep-th/0401162].
  %%CITATION = NUPHA,B703,537;%%

%\cite{Pons:2009ch}
\bibitem{Pons:2009ch}
  J.~M.~Pons,
  ``Substituting fields within the action: consistency issues and some
  applications,''
  arXiv:0909.4151 [hep-th].
  %%CITATION = ARXIV:0909.4151;%%

%\cite{Duff:1985jd}
\bibitem{Duff:1985jd}
  M.~J.~Duff, C.~N.~Pope,
  ``Consistent Truncations In Kaluza-klein Theories,''
  Nucl.\ Phys.\  {\bf B255 } (1985)  355-364.


\bibitem{footnot}The proof is by noticing that 
$\displaystyle \nabla_{\!\rho}\nabla_{\!\alpha}\frac{\partial {\cal L}}{\partial R_{\alpha(\mu\lambda)\rho}}=
\nabla_{\!\rho}\nabla_{\!\alpha}\frac{\partial {\cal L}}{\partial R_{\alpha\mu\lambda\rho}}-
\nabla_{\!\rho}\nabla_{\!\alpha}\frac{\partial {\cal L}}{\partial R_{\alpha[\mu\lambda]\rho}}\,,
$ and that the second term in the rhs is of second order.

%\cite{Lovelock:1971yv}
\bibitem{Lovelock:1971yv}
  D.~Lovelock,
  ``The Einstein tensor and its generalizations,''
  J.\ Math.\ Phys.\  {\bf 12} (1971) 498
  %%CITATION = JMAPA,12,498;%%

%\cite{Mukhopadhyay:2006vu}
\bibitem{Mukhopadhyay:2006vu}
  A.~Mukhopadhyay and T.~Padmanabhan,
  ``Holography of gravitational action functionals,''
  Phys.\ Rev.\  D {\bf 74} (2006) 124023
  [arXiv:hep-th/0608120].
  %%CITATION = PHRVA,D74,124023;%%

%\cite{Yale:2010jy}
\bibitem{Yale:2010jy}
  A.~Yale and T.~Padmanabhan,
  ``Structure of Lanczos-Lovelock Lagrangians in Critical Dimensions,''
  arXiv:1008.5154 [gr-qc].
  %%CITATION = ARXIV:1008.5154;%%

%\cite{Myers:1987yn}
\bibitem{Myers:1987yn}
  R.~C.~Myers,
  ``Higher derivative gravity, surface terms ans string theory,''
  Phys.\ Rev.\  D {\bf 36} (1987) 392.
  %%CITATION = PHRVA,D36,392;%%

%\cite{jorgez2000}
\bibitem{jorgez2000}
R.~Troncoso, J.~Zanelli,
``Higher Dimensional Gravity, Propagating Torsion and AdS Gauge Invariance,''
 Class.\ Quant.\ Grav.\ {\bf 17} (2000) 4451 
[arXiv:hep-th/9907109]

%\cite{Dadhich:2008df}
\bibitem{Dadhich:2008df}
  N.~Dadhich,
  ``Characterization of the Lovelock gravity by Bianchi derivative,''
  Pramana {\bf 74}, (2010) 875 
  [arXiv:0802.3034 [gr-qc]].
  %%CITATION = PRAMC,74,875;%%



\end{thebibliography}
\end{document}